# High elasticity and strength of ultra-thin metallic transition metal dichalcogenides

Ali Sheraz[1], Naveed Mehmood[2], Mert Miraç Çiçek[2,3], İbrahim Ergün[1], Hamid Reza Rasouli[2], Engin Durgun[2], T. Serkan Kasırga[1,2]

[1] Department of Physics, Bilkent University, Ankara, Turkey 06800
[2] Institute of Materials Science and Nanotechnology – UNAM, Bilkent University, Ankara, Turkey 06800
[3] Department of Engineering Physics, Faculty of Engineering, Ankara University, Ankara 06100, Turkey

**Abstract:**

Mechanical properties of transition metal dichalcogenides (TMDCs) are relevant to their prospective applications in flexible electronics. So far, the focus has been on the semiconducting TMDCs, mostly $MoX_2$ and $WX_2$ (X=S, Se) due to their potential in optoelectronics. A comprehensive understanding of the elastic properties of metallic TMDCs is needed to complement the semiconducting TMDCs in flexible optoelectronics. Thus, mechanical testing of metallic TMDCs is pertinent to the realization of the applications. Here, we report on the atomic force microscopy-based nano-indentation measurements on ultra-thin 2H-$TaS_2$ crystals to elucidate the stretching and breaking of the metallic TMDCs. We explored the elastic properties of 2H-$TaS_2$ at different thicknesses ranging from 3.5 nm to 12.6 nm and find that the Young's modulus is independent of the thickness at a value of 85.9 ± 10.6 GPa, which is lower than the semiconducting TMDCs reported so far. We determined the breaking strength as 5.07 ± 0.10 GPa which is 6% of the Young's modulus. This value is comparable to that of other TMDCs. We used *ab initio* calculations to provide an insight to the high elasticity measured in 2H-$TaS_2$. We also performed measurements on a small number of 1T-$TaTe_2$, 3R-$NbS_2$ and 1T-$NbTe_2$ samples and extended our *ab initio* calculations to these materials to gain a deeper understanding on the elastic and breaking properties of metallic TMDCs. This work illustrates that the studied metallic TMDCs are suitable candidates to be used as additives in composites as functional and structural elements and for flexible conductive electronic devices.

**Main Text:**

Two dimensional (2D) layered materials show exceptional mechanical strength along the basal plane direction. Ultra-thin crystals of graphene[1], h-BN[2], $MoS_2$[3–5] and $Ti_3C_2T_x$[6] have record high biaxial Young's modulus with the breaking strength at the intrinsic limit[7]. Such a level of elasticity and strength in these materials is crucial for the applications in the flexible electronics. One particularly appealing family among the 2D layered materials for the prospective applications has been transition metal dichalcogenides (TMDCs). TMDCs are composed of layers formed by covalently bonded transition metals and chalcogen atoms (sulfur, selenium, tellurium) and these layers are stacked via van der Waals interactions. At the monolayer limit, $MoX_2$ and $WX_2$ (X=S, Se) display intriguing optoelectronic properties and attracted a great deal of attention[8]. Among the TMDCs, so far the mechanical properties of ultra-thin $MoS_2$[3–5], $WS_2$[9], $WSe_2$[10], $MoSe_2$[11], and $MoTe_2$[12] has been investigated. Although there are reports on the elastic properties of the bulk samples of other TMDCs, metallic TMDCs are overlooked and the reported elastic properties are not intrinsic to the materials but dominated by the defects at the grain boundaries and pre-existing cracks.

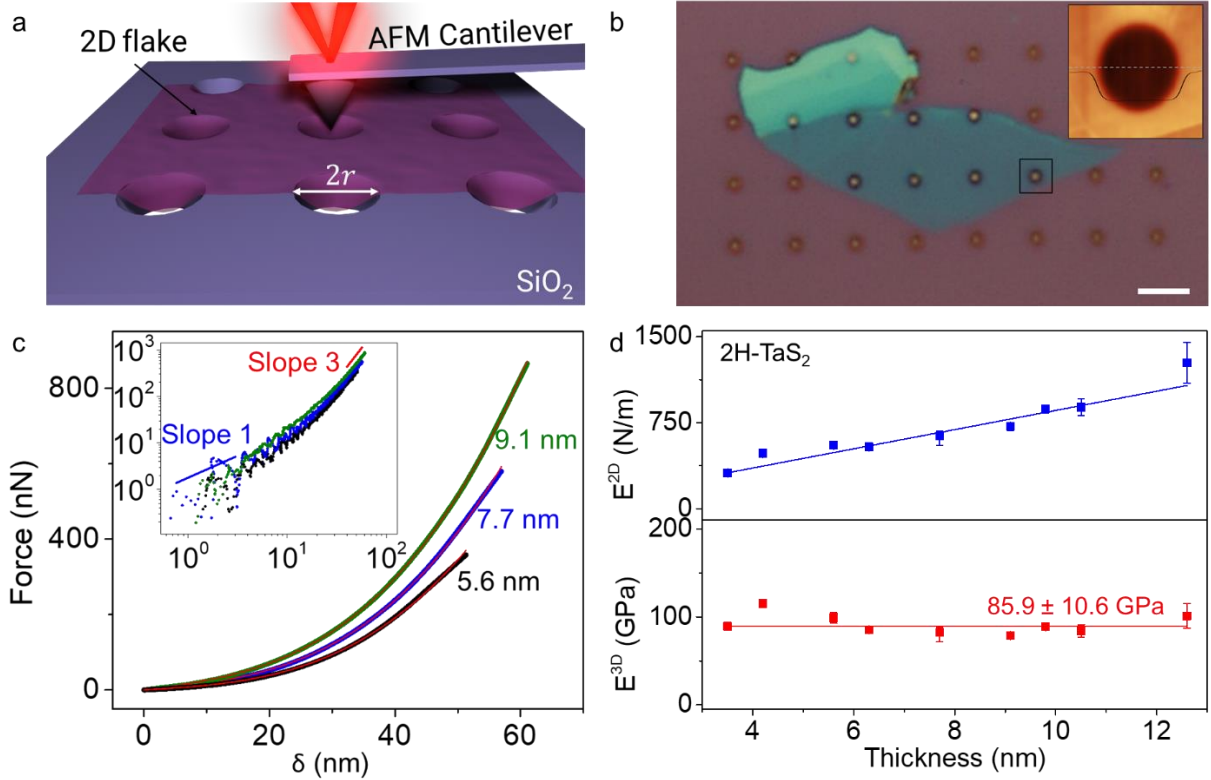

**Figure 1. a.** Schematic of the indentation setup is depicted in the illustration. 2D flake is laid over a hole of radius $r$ etched over the $SiO_2$ surface. **b.** Optical microscope micrograph of a 2H-$TaS_2$ flake transferred over the holes on oxidized Si chip is shown. AFM height map taken from the black square is given in the inset. Thickness of the crystal at the scan area is 6 nm. Height trace shows that the crystal adheres to the sides of the hole. Scale bar is 5 μm. **c.** Force-deflection curves ($F-\delta$) for 2H-$TaS_2$ crystals at different thicknesses, 5.6 nm, 7.7 nm and 9.1 nm. The colored curves are the experimental data and the red solid lines are fit to eq.(1). Inset shows the log-log plot with linear response of the crystal in the first few nanometers of indentation and approaches to the cubic response at the higher loads. **d.** 2D elastic moduli ($E^{2D}$) and Young's moduli ($E^{3D}$) for $TaS_2$ at different thickness are given in the plot. Each data point is determined from multiple measurements from a total of 26 different crystals. The average Young's modulus is $85.9 \pm 10.6$ GPa as denoted with the red line through the graph. $E^{2D}$ increases linearly with the thickness similar to other TMDCs.

Here, we elucidated the elastic properties of 2H-$TaS_2$, a prototypical metallic TMDC, using atomic force microscopy (AFM) based nanoindentation and supported our findings with *ab initio* calculations. We performed comprehensive measurements on 5 to 19 monolayers of 2H-$TaS_2$. We also performed nanoindentation on a small number of 1T-$TaTe_2$, 1T-$NbTe_2$ and 3R-$NbS_2$ crystals and performed *ab initio* calculations. All the materials except 3R-$NbS_2$ are exfoliated from the bulk crystals in the ambient using a sticky tape over a PDMS stamp and transferred over circular holes drilled using focused ion beam (FIB). We used chemical vapor deposition (CVD) method to grow thin layers of 3R-$NbS_2$ and[13] transferred them over the holes using a wet transfer method (details are provided in the Supporting Information, **Figure S1-S4**). 3R-polytype of $NbS_2$ has the same crystal structure as the monolayer but different stacking in the multilayer. The details of the exfoliation and the CVD synthesis are given in the Supporting Information. The polytype of each material is confirmed using Raman spectroscopy[14,15], shown in **Figure S5**. Optical microscope micrographs of exemplary crystals over the indentation substrates are shown in **Figure S6.**

First, we would like to focus on our measurements on 2H-TaS$_2$. 2H-TaS$_2$ shows a thickness dependent enhancement in the superconducting transition temperature ($T_C \approx 0.5$ K) and becomes a 2D superconductor below ~10 nm[16]. It is well known that both the charge density wave transition and the superconductivity in bulk 2H-TaS$_2$ is effected by hydrostatic pressure on the crystal[17]. Moreover, with a 5.6 eV work function, it is a suitable candidate for rectifying contacts to semiconducting TMDCs[18]. Thus, it is imperative to understand the elastic properties of metallic 2H-TaS$_2$ for possible device applications in low temperature and flexible electronics. 2H-TaS$_2$ is relatively difficult to exfoliate and the thinnest crystal we could measure was 5 monolayers thick (with ~0.7 nm monolayer thickness). 26 samples of 2H-TaS$_2$ are measured in total with thicknesses ranging from 3.5 nm to 12.6 nm. The crystal thicknesses are determined with respect to the supporting substrate surface via AFM scan.

Schematic of the measurement setup is given in **Figure 1a**. Diameter of the holes etched on the SiO$_2$ surface for suspending the ultra-thin crystals is measured as $r = 1$ μm using scanning electron microscope (SEM), which is consistent with the radius determined from the AFM measurements. Optical microscope image of a TaS$_2$ crystal transferred over the substrate is shown in **Figure 1b**. The AFM cantilevers used for the indentation studies have the spring constant $k = 40$ N/m and the tip radius $r_{tip} = 10$ nm with diamond like coating, specified by the manufacturer. We preferred a stiffer cantilever to be able to indent the crystals as the thinnest crystal we could find was 5 monolayers. The details regarding the calibration of the AFM cantilevers' spring constant and deflection sensitivity are given in the Supporting Information and **Figure S7** [19]. To preserve the sample properties, we limited the device fabrication duration to a few minutes, and we finished the indentation measurements within an hour after the crystal transfer. However, we did not observe any statistically significant difference between the Young's moduli obtained from the shorter and the longer duration measurements for 2H-TaS$_2$. Dummy samples are scanned for at least half an hour before the indentation measurements to minimize the piezo drift in the tapping mode. Then, the actual sample is scanned, and the tip is positioned at the center of the chosen hole for the indentation studies. We applied gradually increasing forces on the crystals till the brittle fracture.

The force vs. displacement ($F - \delta$) curves obtained from the indentation measurements are fitted using the formula for an elastic membrane under nonlinear deformation[1,4]:

$$F = \left[\frac{4\pi E^{2D} t^2}{3(1-v^2)r^2}\right]\delta + \sigma_0^{2D}\pi\delta + E^{2D}\frac{q^3}{r^2}\delta^3 \qquad (1)$$

where $E^{2D}$ is the 2D elastic modulus, $\sigma_0^{2D}$ is the prestress in the membrane, $r$ is the radius of the hole, $q$ is a dimensionless constant related to the Poisson's ratio $v$ as $q = 1/(1.049 - 0.15v - 0.16v^2)$ and $t$ is the thickness of the crystal. Here, the first term represents the stiffness of a plate with a certain bending rigidity and becomes significant for thicker crystals. In our case, as the thickness of the crystals are considerably smaller than the radius of the hole, it is ignorable. The second term is the pretension of a stretched membrane under force and the last term is the nonlinear membrane behavior. The last term is dominant at large loads. Before fitting the measurement data with equation (1), we determined the point where both the force and displacement are zero by intersecting the line extrapolated from the zero-force line in the $F-\delta$ curve with the curved section[6]. We checked some of our results with a revised formula that eliminates the need for determination of the zero force and displacement point (ZDP) by fitting the zero force and the membrane displacement parameters[20]. We obtained identical results with both methods within the error margins. Very large coefficient of determination, $R^2 > 0.998$ are achieved in fitting all the measurements with eq.(1). **Figure 1c** shows the $F-\delta$ curve for crystals with three different thicknesses and the fit of the experimental data with eq.(1). Repeated loading-unloading curves follow the same $F-\delta$ curves show that there is no slippage of the crystals over the holes, even for the ones near the edge.

We measured that 2H-TaS$_2$ has a thickness independent Young's modulus value of $85.9 \pm 10.6$ GPa with the Poisson's ratio of $\nu_{TaS_2} = 0.27$ determined by *ab initio* calculations. Young's modulus values for different thicknesses are given in **Figure 1d**. To best of our knowledge, the room temperature Young's modulus for 2H-TaS$_2$ has not been reported in the literature experimentally. Based on the velocity of ultrasonic waves measured in 2H-TaS$_2$ below 65 K [21], we estimated the Young's modulus of 2H-TaS$_2$ as 82 GPa as $E^{3D} = v_p^2 \rho (1 - \nu^2)$ where $\rho = 6.86$ g/cm$^3$ is the density of 2H-TaS$_2$ and $v_p$ is the velocity of the quasi-longitudinal wave that has the highest propagation velocity. Thickness independence of Young's modulus from bulk down to a few layers can be explained by large shear strain energy of TaS$_2$ layers[2]. Another parameter we can extract from the equation 1 by fitting the indentation data is the pretension $\sigma_0^{2D}$ of the suspended membrane. The values range from 1.2 N/m to 10.2 N/m in our case and the pretension is in an increasing trend with the crystal thickness (**Figure S8**). This shows a strong interaction between the membrane and the hole walls.

We also determined the breaking stress, $\sigma_m^{2D}$, and the breaking strength, $\sigma_m$, using the equation $\sigma_m^{2D} = \sqrt{\frac{F_m E^{2D}}{4\pi r_{tip}}}$, derived based on the continuum model of a circular membrane in the elastic regime under a spherical indenter with radius $r_{tip}$ and breaking force of $F_m$ [1]. Loading and unloading of the membrane virtually repeating the same $F-\delta$ curve shows that the membranes are in the elastic regime (**Figure 2a**). Here, the model assumes that $\frac{r_{tip}}{r} \ll 1$. However, measuring the $r_{tip}$ is imperative due to our choice of stiff cantilever. We carefully characterized the AFM tips using SEM to ensure that we use the correct $r_{tip}$ value for calculating the breaking strength. First, we measured the pristine tip radius, that turns out to be within the specifications of the manufacturer for all the tips we used. Then, we measured the tip radius after a scan over the crystal and after an indentation without breaking. This $r_{tip}$ value is used for the breaking strength calculation. Side-by-side comparison of the SEM micrographs of the tips (**Figure 2b)** show that the tips are slightly worn after each experiment to a certain extent. Some tips become significantly blunt after breaking the crystal, due to sudden strike of the tip to the base of the hole (**Figure S9**). Thus, we used each tip for a single breaking (and indentation) measurement. **Figure S10** shows a series of AFM tips used in scanning, indentation, and breaking experiments. We found that the tip radius is typically within $r_{tip} = 22 \pm 2$ nm after each experiment. $\sigma_m^{2D}$ increases with the increasing crystal thickness with an average thickness independent breaking strength of $5.07 \pm 0.10$ GPa as shown in **Figure 2c**. The breaking strength of 2H-TaS$_2$ crystals are 6% of their Young's modulus. This is very close to the breaking strength/Young's modulus ratio that is reported for the few layer MoS$_2$[3].

The breaking strength is comparable to the monolayer MoS$_2$ and WSe$_2$ making 2H-TaS$_2$ a considerable choice to form Schottky junctions for flexible electronics[18].

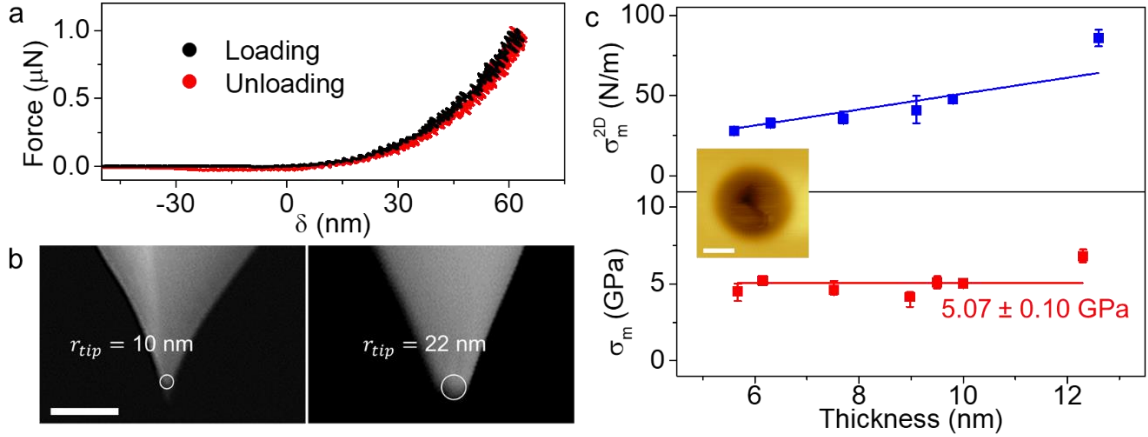

**Figure 2. a.** Loading-Unloading curve of the same crystal shows the elastic behavior of the membrane. ZDP is not corrected. **b.** SEM images of the AFM tip before and after the indentation show that the tip radius almost doubles due to the wear. Scale bar is 100 nm. **c.** Upper panel shows the 2D breaking strength of 2H-TaS$_2$ membranes at different thicknesses with the AFM height map of a fractured crystal in the inset. Scale bar is 500 nm. Lower panel shows the ultimate strength of the material with an average of $5.07 \pm 0.10$ GPa.

We performed *ab initio* calculations to gain a deeper understanding of our experimental results as well as to provide a validation to our measurements (methods are provided in the supporting information[23–31]). We also performed experiments on three other thin crystals; 3R-NbS$_2$, 1T-NbTe$_2$ and 1T-TaTe$_2$ to observe the general trend in metallic TMDCs. **Figure 3a** and **3b** show the crystal structures of the monolayers of various metallic TMDCs. All the crystal structures and crystal parameters are given in **Figure S11**. The breaking strength values are calculated via the strain vs. stress curves (**Figure 3c**) which are obtained by stretching the monolayer crystals in biaxial direction. The deviation from the experimental data is possibly obtained due to the stacking fault, crystal orientation, and defect density which are not included in the calculations[10]. The strain-stress curves are obtained by applying the strain up to 25%.

To compare the brittleness of the metallic TMDCs to other 2D layered materials, we calculated the surface energy, $\epsilon_{BP}$, in the basal plane. The edge energy of a monolayer is defined as $\epsilon_{edge} = \frac{(E_{NR} - E_{Mono})}{2L}$, where $E_{NR}$ and $E_{Mono}$ are the total energy of the nanoribbon and the pristine monolayer (with the same number of atoms), respectively and, *L* is the ribbon length. In these calculations, the zigzag- and armchair-edged nanoribbons were constructed with 24 and 36 atoms, respectively. The surface energy was determined by the relation, $\epsilon_{BP} = \epsilon_{edge} t$, where *t* is the thickness corresponding to the interplanar spacing. The model and the obtained results agree with the literature[11,32]. Based on the lowest surface energy of the basal plane, $\epsilon_{BP}$, we can estimate the fracture toughness, $K_{FT}$, of the materials. Here, we define the fracture toughness as the critical strain energy release rate of the fracture, as common in the literature[7]. For brittle materials, the fracture toughness is roughly twice the surface energy of the basal plane[11]. We obtained 1.14, 1.11, 0.39, and 0.35 Jm$^{-2}$ for the lowest surface energy of the basal plane of 2H-TaS$_2$, 3R-NbS$_2$, TaTe$_2$, NbTe$_2$ respectively. **Figure 3d** shows $K_{FT}$ vs. $\sigma_m/E^{3D}$ for various 2D layered materials. An ideal material for high strength applications should have large ultimate strain and fracture toughness. Graphene shows exceptional mechanical properties whereas $MoS_2$ and $MoSe_2$ has higher fracture toughness than all four materials we studied in this work with lower $\sigma_m/E^{3D}$ ratio. These insights are relevant for engineering applications of these materials.

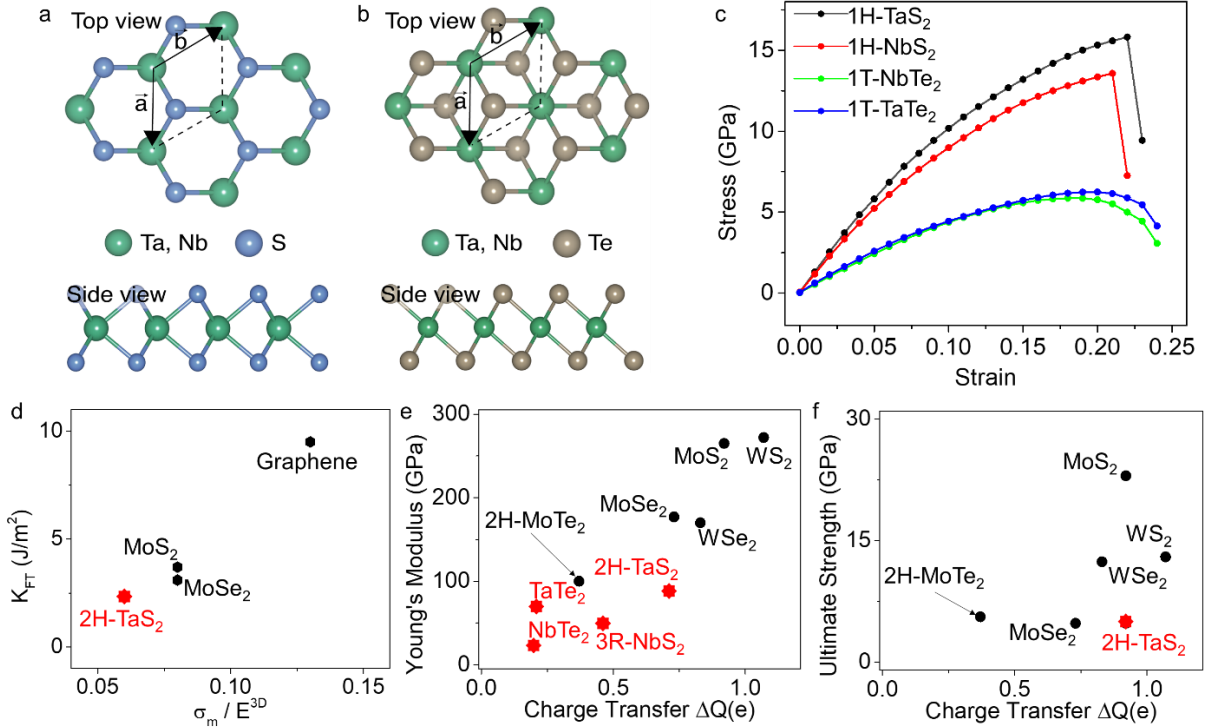

**Figure 3 a.** and **b.** show the top and side views of the 1H (sulfides) and 1T (tellurides) monolayers of the materials studied in this work. **c.** Biaxial stress-strain curve calculated using DFT till the instability of the lattice. **d.** Fracture toughness ($K_{FT}$) vs. breaking strength/Young's modulus ($\sigma_m/E^{3D}$) ratio for various 2D layered materials. Graphene shows the highest fracture toughness with remarkably high $\sigma_m/E^{3D}$ ratio while the 2H-$TaS_2$ we measured in this study exhibit a lower value in trend with other materials[1,11,33]. **e.** Experimentally determined Young's modulus vs. charge transfer is plotted in the figure for the metallic TMDCs. Charge transfer values are obtained via *ab initio* calculations for monolayer crystals. As the charge transfer from transition metal to chalcogen becomes larger, Young's modulus increases. **f.** A similar trend is observed in the ultimate strength of the materials[3,9–12,34]. Data point marked by red star represents the results from this work.

We investigated the bulk charge transfer in TMDCs studied here via DFT and compared the results to Mo and W based TMDCs. Mechanical properties of TMDCs are strongly influenced by the bonding charge distribution[34]. DFT studies by Li et al.[34] show that for the TMDCs, the Young's modulus and the ultimate strength of sulfides are the largest and selenides are larger than the tellurides with the same transition metal. The charge density between the chalcogen and the transition metal atoms decreases from sulfides to tellurides. This results in weakening of the covalent bonding in the basal plane. Our measurements on 2H- and 1T-crystals confirm the weakening of the covalent bonding in the basal plane. The charge transfer values ($\Delta Q$) are correlated with the Young's moduli of these materials except 3R-crystal due to different stacking orders. **Figure 3e** and **3f** show the measured Young's moduli and the ultimate strength of various 2D layered materials versus the charge transfer values from DFT, respectively.

**Table 1.** Measured and calculated parameters for the metallic TMDCs at room temperature. DFT results are given in italic.

|  | **2H-TaS$_2$** | **1T-TaTe$_2$** | **3R-NbS$_2$** | **1T-NbTe$_2$** |
|---|---|---|---|---|
| **Young's Modulus (GPa) ($E^{3D}$)** | 85.9 ± 10.6 | 70 ± 14 | 49.4 ± 3.0 | 23.6 ± 1.6 |
|  | *77.7* | *70.9* | *56.2* | *67.6* |
| **Breaking Strength (GPa) ($\sigma_m$)** | 5.01 ± 0.10 | 7.18 ± 0.40 | 5.0 ± 1.5 | 2.9 ± 0.3 |
|  | *15.8* | *6.2* | *13.6* | *5.9* |
| **Breaking Strength / Young's modulus** | 0.06 ± 0.01 | 0.10 ± 0.03 | 0.10 ± 0.04 | 0.12 ± 0.02 |
| **Poisson's Ratio ($v$)** | *0.27* | *0.10* | *0.27* | *0.11* |
| **Number of samples measured** | 26 | 4 | 2 | 7 |

All the measured and calculated parameters for the metallic TMDCs studied here are listed in **Table 1.** The deviations from the experimental data, particularly for NbTe$_2$, can be explained by the rapid oxidation of the NbTe$_2$ surface leading to a decrease in the measured Young's modulus and breaking strength value[35]. To test this hypothesis, we measured the oxidation of TaS$_2$, TaTe$_2$, NbTe$_2$ and NbS$_2$ surfaces using XPS. Freshly cleaved crystals of all materials show no signs oxidation while after one-hour in the ambient all materials except TaS$_2$ show new peaks in both the metal and the oxygen XPS surveys that can be attributed to formation of metal-oxides (**Figure S12**). This measurement shows that metallic TMDCs oxidize faster under the ambient and extra care should be taken to prevent oxidation of samples during the indentation studies. Details of the XPS surveys are given in the supporting information. Rate of oxidation possibly plays a role in the measured values of the Young's modulus and the breaking strength. Rapidly oxidizing crystals show more significant deviations from the measured values.

In conclusion, we reported the elastic and breaking properties of 2H-TaS$_2$ and provided an insight into other layered Ta and Nb based metallic TMDCs. Our findings show that these materials can sustain large strains similar to semiconducting 2D materials and can be used in conjunction with other TMDCs as the electrical contact materials for flexible electronics and optoelectronics. The metallic TMDCs can endure similar maximum strain as the semiconducting TMDCs. All the materials we studied in this work are possible candidates as contact materials to the semiconducting TMDCs. Moreover, their strain dependent properties can be exploited in various applications via strain engineering.

**Author Contributions**

TSK proposed and conducted the experiments. AS performed the indentation studies, prepared the samples and the substrates. NM prepared the samples, performed measurements on the crystals and aided the data analysis. IE performed the data analysis with the help of AS. HRR performed the XPS measurements. MMC and ED performed the *ab initio* studies. All authors contributed to the writing of the paper and discussed the results.

**Acknowledgements**

TSK gratefully acknowledges the support from TUBITAK under grant no 118F061. We would like to thank Aizimaiti Aikebeier for discussions on sample preparation.

**Availability of Data**

The data that support the findings of this study are available from the corresponding author upon reasonable request.

**References**

[1] C. Lee, X. Wei, J.W. Kysar, and J. Hone, Science (80-. ). **321**, 385 (2008).

[2] A. Falin, Q. Cai, E.J.G. Santos, D. Scullion, D. Qian, R. Zhang, Z. Yang, S. Huang, K. Watanabe, T. Taniguchi, M.R. Barnett, Y. Chen, R.S. Ruoff, and L.H. Li, Nat. Commun. **8**, 15815 (2017).

[3] S. Bertolazzi, J. Brivio, and A. Kis, ACS Nano **5**, 9703 (2011).

[4] A. Castellanos-Gomez, M. Poot, G.A. Steele, H.S.J. van der Zant, N. Agraït, and G. Rubio-Bollinger, Adv. Mater. **24**, 772 (2012).

[5] Y. Li, C. Yu, Y. Gan, P. Jiang, J. Yu, Y. Ou, D.-F. Zou, C. Huang, J. Wang, T. Jia, Q. Luo, X.-F. Yu, H. Zhao, C.-F. Gao, and J. Li, Npj Comput. Mater. **4**, 49 (2018).

[6] A. Lipatov, H. Lu, M. Alhabeb, B. Anasori, A. Gruverman, Y. Gogotsi, and A. Sinitskii, Sci. Adv. **4**, 1 (2018).

[7] A.A. Griffith, Philos. Trans. R. Soc. London. Ser. A, Contain. Pap. a Math. or Phys. Character **221**, 163 (1921).

[8] G. Wang, A. Chernikov, M.M. Glazov, T.F. Heinz, X. Marie, T. Amand, and B. Urbaszek, Rev. Mod. Phys. **90**, 021001 (2018).

[9] K. Liu, Q. Yan, M. Chen, W. Fan, Y. Sun, J. Suh, D. Fu, S. Lee, J. Zhou, S. Tongay, J. Ji, J.B. Neaton, and J. Wu, Nano Lett. **14**, 5097 (2014).

[10] R. Zhang, V. Koutsos, and R. Cheung, Appl. Phys. Lett. **108**, 042104 (2016).

[11] Y. Yang, X. Li, M. Wen, E. Hacopian, W. Chen, Y. Gong, J. Zhang, B. Li, W. Zhou, P.M. Ajayan, Q. Chen, T. Zhu, and J. Lou, Adv. Mater. **29**, 1604201 (2017).

[12] Y. Sun, J. Pan, Z. Zhang, K. Zhang, J. Liang, W. Wang, Z. Yuan, Y. Hao, B. Wang, J. Wang, Y. Wu, J. Zheng, L. Jiao, S. Zhou, K. Liiu, C. Cheng, W. Duan, Y. Xu, Q. Yan, and K. Liu, Nano Lett. (2019).

[13] X. Wang, J. Lin, Y. Zhu, C. Luo, K. Suenaga, C. Cai, and L. Xie, Nanoscale **9**, 16607 (2017).

[14] S. Nakashima, Y. Tokuda, A. Mitsuishi, R. Aoki, and Y. Hamaue, Solid State Commun. **42**, 601 (1982).

[15] M. Hangyo, S.-I. Nakashima, and A. Mitsuishi, Ferroelectrics **52**, 151 (1983).

[16] E. Navarro-Moratalla, J.O. Island, S. Manãs-Valero, E. Pinilla-Cienfuegos, A. Castellanos-Gomez, J. Quereda, G. Rubio-Bollinger, L. Chirolli, J.A. Silva-Guillén, N. Agraït, G.A. Steele, F. Guinea, H.S.J. Van Der Zant, and E. Coronado, Nat. Commun. **7**, 1 (2016).

[17] D.C. Freitas, P. Rodière, M.R. Osorio, E. Navarro-Moratalla, N.M. Nemes, V.G. Tissen, L. Cario, E. Coronado, M. García-Hernández, S. Vieira, M. Núñez-Regueiro, and H. Suderow, Phys. Rev. B **93**, 184512 (2016).

[18] T. Shimada, F.S. Ohuchi, and B.A. Parkinson, Jpn. J. Appl. Phys. **33**, 2696 (1994).

[19] I.W. Frank, D.M. Tanenbaum, A.M. van der Zande, and P.L. McEuen, J. Vac. Sci. Technol. B Microelectron. Nanom. Struct. **25**, 2558 (2007).

[20] Q.Y. Lin, G. Jing, Y.B. Zhou, Y.F. Wang, J. Meng, Y.Q. Bie, D.P. Yu, and Z.M. Liao, ACS Nano **7**, 1171 (2013).

[21] M.H. Jericho, A.M. Simpson, and R.F. Frindt, Phys. Rev. B **22**, 4907 (1980).

[22] J.W. Suk, R.D. Piner, J. An, and R.S. Ruoff, ACS Nano **4**, 6557 (2010).

[23] P. Hohenberg and W. Kohn, Phys. Rev. **136**, B864 (1964).

[24] W. Kohn and L.J. Sham, Phys. Rev. **140**, A1133 (1965).

[25] G. Kresse and J. Furthmüller, Comput. Mater. Sci. **6**, 15 (1996).

[26] P.E. Blöchl, Phys. Rev. B **50**, 17953 (1994).

[27] J.P. Perdew, K. Burke, and M. Ernzerhof, Phys. Rev. Lett. **77**, 3865 (1996).

[28] G. Henkelman, A. Arnaldsson, and H. Jónsson, Comput. Mater. Sci. **36**, 354 (2006).

[29] D. Kecik, A. Onen, M. Konuk, E. Gürbüz, F. Ersan, S. Cahangirov, E. Aktürk, E. Durgun, and S. Ciraci, Appl. Phys. Rev. **5**, 011105 (2018).

[30] S. Singh, I. Valencia-Jaime, O. Pavlic, and A.H. Romero, Phys. Rev. B **97**, 054108 (2018).

[31] R. Hill, Proc. Phys. Soc. Sect. A **65**, 349 (1952).

[32] P. Koskinen, S. Malola, and H. Häkkinen, Phys. Rev. Lett. **101**, 115502 (2008).

[33] P. Zhang, L. Ma, F. Fan, Z. Zeng, C. Peng, P.E. Loya, Z. Liu, Y. Gong, J. Zhang, X. Zhang, P.M. Ajayan, T. Zhu, and J. Lou, Nat. Commun. **5**, 3782 (2014).

[34] J. Li, N. V. Medhekar, and V.B. Shenoy, J. Phys. Chem. C **117**, 15842 (2013).

[35] R. Yan, G. Khalsa, B.T. Schaefer, A. Jarjour, S. Rouvimov, K.C. Nowack, H.G. Xing, and D. Jena, Appl. Phys. Express **12**, 023008 (2019).

# Supporting Information: High elasticity and strength of ultra-thin metallic transition metal dichalcogenides

Ali Sheraz[1], Naveed Mehmood[2], Mert Miraç Çiçek[2,3], İbrahim Ergün[1], Hamid Reza Rasouli[2], Engin Durgun[2], T. Serkan Kasırga[1,2]

[1] Department of Physics, Bilkent University, Ankara, Turkey 06800
[2] Institute of Materials Science and Nanotechnology – UNAM, Bilkent University, Ankara, Turkey 06800
[3] Department of Engineering Physics, Faculty of Engineering, Ankara University, Ankara 06100, Turkey

**Methods**

We obtained bulk crystals from HQ Graphene and exfoliated using Nitto blue tape over polydimethylsiloxane (PDMS) stamps for the transfer. Then, using a deterministic dry transfer method, thin crystals are transferred over the holes drilled on $SiO_2$ substrate. 3R-$NbS_2$ crystals are synthesized on sapphire substrate via salt-assisted chemical vapor deposition in a split-tube furnace. The growth recipe is provided in the following section. MFP-3D Asylum Research AFM is used for the indentation experiments. We used Tap300Al-G and Tap300DLC diamond-like coated tips from the Budget Sensors with force constant of 40 N/m. As mentioned in the main text, a dummy sample is scanned for about an hour before we conduct the indentation experiments on the real sample to limit the exposure of our samples to the ambient. Further scans are performed over the actual sample till the signs of drift disappears. During the indentation, $Z_{Piezo}$ displacement speed was controlled at 100 nm/s to get hysteresis free loading and unloading curves. We have chosen to wrinkle free flakes with large surface area, uniformly suspended over the holes to obviate inaccuracy, hysteresis, and slippage of flakes.

Deflection ($\delta$) of the suspended flake can be found from the deflection of AFM cantilever ($\Delta Z_C$) and the scanning piezotube displacement ($\Delta Z_{Piezo}$) through[1]:

$$\delta = \Delta Z_{Piezo} - \Delta Z_C$$

Force applied to suspended 2D flakes can be calculated using Hook`s law $F = K_C \Delta Z_C$ here $K_C$ is deflection of AFM cantilever (40 Nm$^{-1}$) that we calibrated using GetReal calibration in Asylum AFM, which first performs the thermal noise spectrum measurement followed by the Sader's method calibration. Finally, the thermal noise method is used one more time to obtain the inverse optical lever sensitivity. This method is slightly better than the hard strike in preserving the cantilever tip.

First-principles calculations based on density functional theory (DFT)[2,3] as implemented in the Vienna Ab Initio Simulation Package (VASP)[4] were performed. The Kohn-Sham equations were solved using the projector augmented-wave method[5] and the exchange-correlation functional was described within the generalized gradient approximation (GGA)[6]. The kinetic-energy cutoff for plane wave basis set taken to be 550 eV and 15x15x1, 15x15x11, 9x15x11 Gamma-centered k-point mesh was used for the numerical integrations over the Brillouin zone for 2H-, 3R-, 1T-crystals, respectively. The lattice constants and atomic positions were optimized until the total energy and force convergence were below 10$^{-6}$ eV and 0.01 eV Å$^{-1}$, respectively. A vacuum space of 20 Å was taken along non-period direction to avoid the interactions between the neighboring images for monolayer crystals. The charge transfer analysis was performed by using Bader technique[7]. The mechanical response of monolayers in the elastic regime was determined by calculating the in-plane stiffness, $Y_{2D} = \frac{c_{11}^2 - c_{12}^2}{c_{11}^2}$, where the $c_{ij}$'s are the elastic constants including hydrostatic and shear terms[8]. The Young modulus of bulk crystals were calculated by using open-source code[9] based on Voigt-Reuss-Hill averaging scheme[10]. A denser k-point mesh was taken for calculations on mechanical properties.

**CVD Growth of $NbS_2$ and Wet Transfer Method**

$NbS_2$ flakes were grown using ambient pressure chemical vapor deposition method on a c-cut sapphire substrate. Before the growth, the sapphire substrates are cleaned using acetone, isopropanol, water and dried by blowing $N_2$ gas. Niobium pentoxide ($Nb_2O_5$) powder, crushed salt (NaCl) and Sulfur are used as the growth precursors. $Nb_2O_5$ and NaCl are milled together using mortar and pestle. The mixture is transferred in an alumina boat and the sapphire substrate is placed over the boat facing down at a few millimeters to the precursor powder. Sulphur precursor is placed in a different alumina boat. Both precursors are positioned in a quartz tube and the tube is slid into the tube furnace. **Figure S1** shows the schematic of the growth setup.

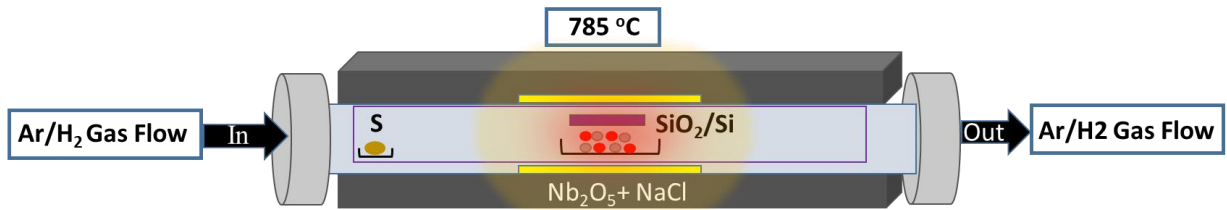

**Figure S1** Schematic of CVD growth chamber for 3R-$NbS_2$ crystals showing different temperature zones and configuration of sample and precursors.

Once the chamber is sealed, we purged the chamber using Ar gas flow rate 499 sccm for 6 minutes to get rid of the air inside the chamber. After 6 minutes, we set Ar at 20 sccm flow rate and started the CVD chamber heater then waited for it to reach at 785 °C. After 785 °C is achieved, Ar and $H_2$ flow 100 sccm and 12 sccm respectively. After 10-15 minutes, we stopped the $H_2$ flow and set Ar again at 20 sccm during cool down. We got large area thin $NbS_2$ flakes which are exceedingly difficult to obtain via mechanical exfoliation. **Figure S2** shows some examples of the 3R-$NbS_2$ crystals on sapphire substrate.

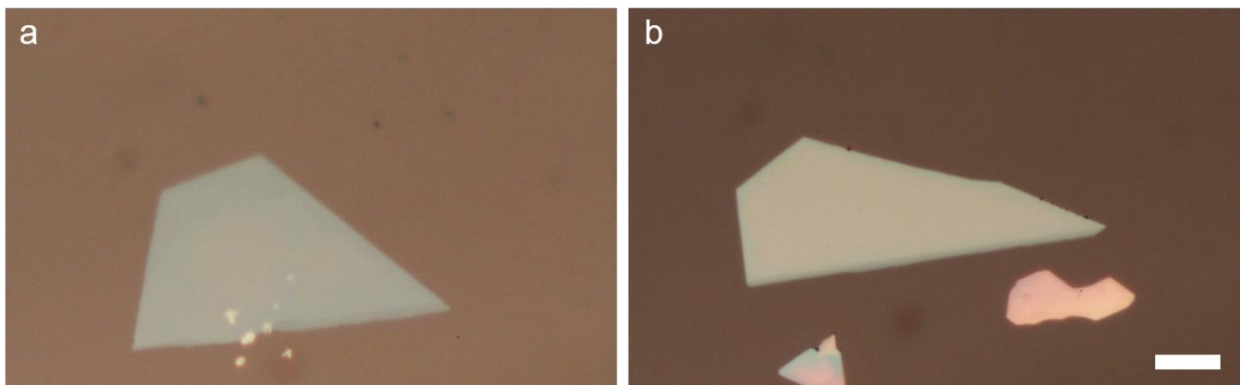

**Figure S2 a** and **b** show the optical microscope micrograph of 3R-NbS$_2$ thin crystals grown on c-cut sapphire via CVD. Scale bar is 10 μm.

XPS and EDX maps taken from the NbS$_2$ flakes show X-Ray Na intercalation within the crystals. **Figure S3** shows the XPS surveys for Nb, S and, Na after various etching cycles. We hypothesize that there might be Na intercalation between the layers as there are reports in the literature showing Na-NbS$_2$ crystals.

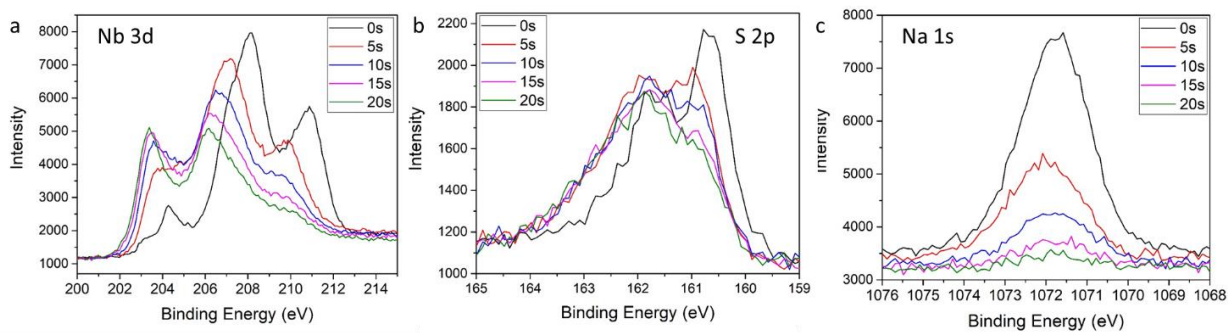

**Figure S3** XPS spectra for **a.** Na 3d, **b.** S 2p and, **c.** Na 1s peaks after various etching cycles.

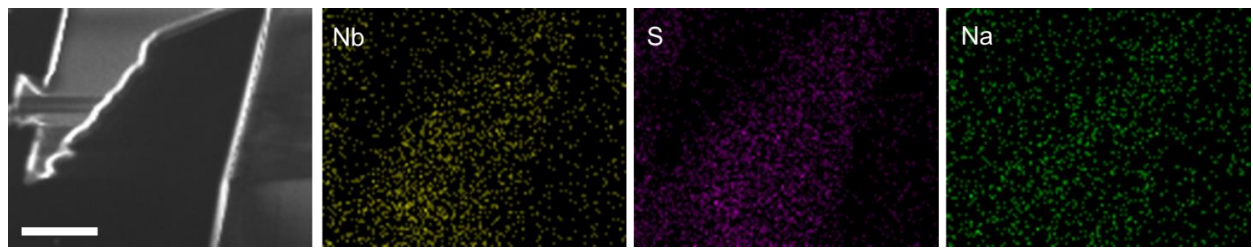

**Figure S4** SEM image and corresponding EDX maps show the existence of Nb, S and Na in the crystal.

**Wet Transfer of NbS$_2$ Crystals**

Sapphire substrate with CVD grown NbS$_2$ is spin coated with Poly(methyl methacrylate) PMMA 495 (A4) at 1300rpm. The substrate is heated at 180 °C for 6 minutes. Then, the substrate is dipped into buffered oxide etch (BOE) for 20 minutes to release the PMMA film along with the CVD grown crystals from the substrate. After rinsing, the film is released onto water surface by wedge transfer method and subsequently picked up by PDMS stamp. The PMMA film is dissolved in acetone for 20 minutes to transfer $NbS_2$ crystals successfully. Finally, the desired crystal was transferred onto the holes using deterministic dry transfer technique as mentioned in main text.

**Raman Spectra of $2H\text{-}TaS_2$, $3R\text{-}NbS_2$, $1T\text{-}TaTe_2$, $1T\text{-}NbTe_2$**

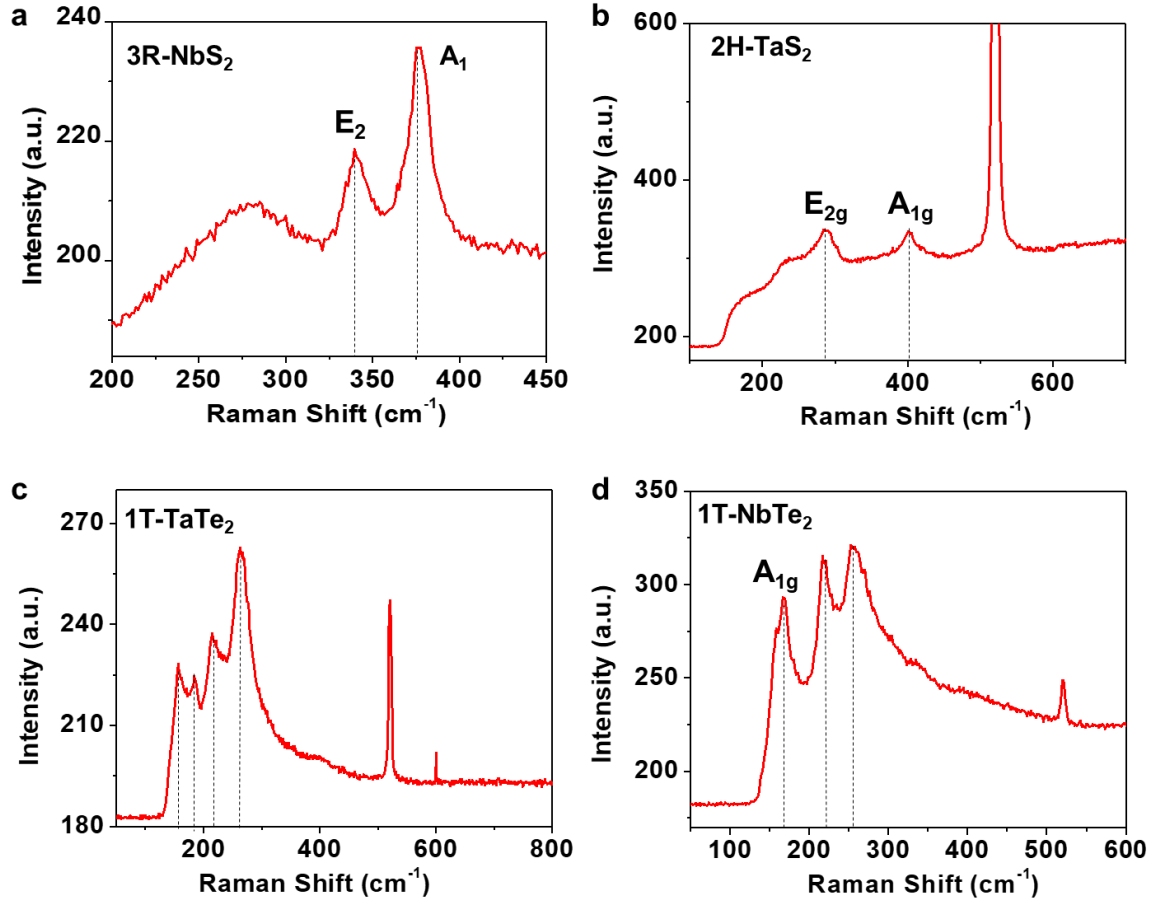

**Figure S5** Raman spectra taken from the crystals studied in this work. **a.** Raman spectrum of $3R\text{-}NbS_2$ **b.** $2H\text{-}TaS_2$ **c.** $TaTe_2$ and **d.** $NbTe_2$. Identified modes are labelled on the corresponding peaks. Dashed vertical lines show the matching peaks with the literature.

We performed Raman spectroscopy to confirm the phase of the crystals studied in this work. **Figure S5** shows the exemplary Raman spectra we obtained for each crystal. Dashed lines indicated on the spectra shows the matching lines with the literature[11,12]. Although the Raman spectrum of $NbS_2$ matches with the spectrum reported for $3R\text{-}NbS_2$ in the literature, as we discuss in the following sections, X-ray photoelectron spectroscopy (XPS) and energy dispersive spectroscopy (XPS) shows that the NbS$_2$ crystals we have might have Na atoms interstitially between the layers.

**Exemplary Optical Microscope Micrographs of Suspended Crystals**

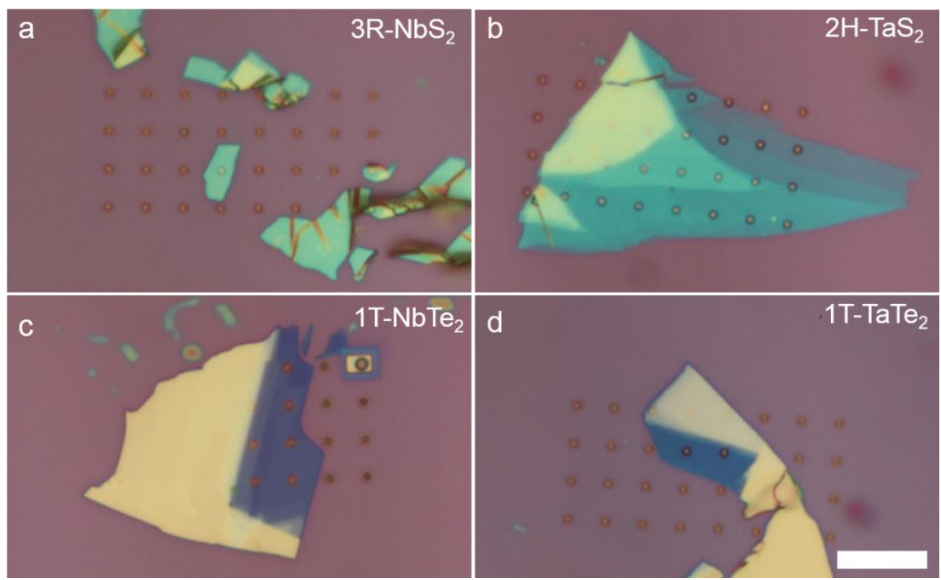

**Figure S6** Optical microscope micrograph of **a.** 3R-NbS$_2$ **b.** 2H-TaS$_2$ **c.** 1T-NbTe$_2$ and **d.** 1T-TaTe$_2$. Scale bar is 10 µm.

**Calibration of AFM Probes using GetReal Calibration Method**

We used built-in GetReal method for automated calibration of AFM probes that uses Sader's method[13] and thermal noise method for the calibration of cantilever's spring constant and sensitivity (InvOLS) respectively.

**Force-Indentation Loading Curve**

A full cycle of the $F - \delta$ loading curve is given in **Figure S7**. Important points in the curve are marked on the graph. First the tip starts above the membrane. When the tip is sufficiently close, the tip snaps onto the membrane and indentation continues till the mechanical failure of the film. Then, the tip rapidly descends to the bottom of the hole and starts applying force there again.

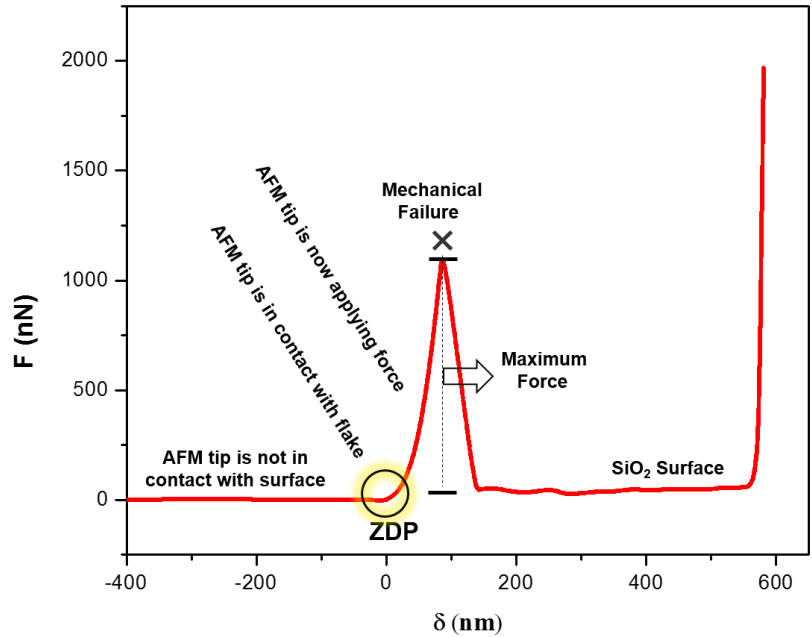

**Figure S7.** Figure shows the complete $F - \delta$ graph for a typical crystal reported in this work. Upon lowering the tip in the close vicinity of the suspended membrane, the membrane snaps to the tip around the zero-displacement point (ZPD). Then, further lowering the tip stretches the membrane till the brittle mechanical failure. Then, the tip reaches down to the bottom of the hole till it touches the SiO$_2$ surface.

**Pretension and Prestress of 2H-TaS$_2$**

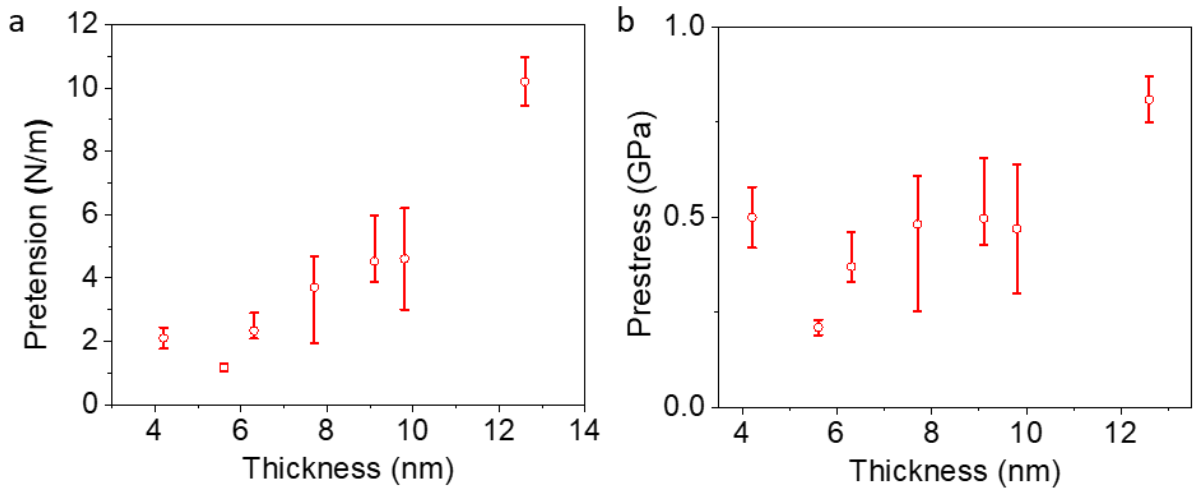

**Figure S8 a.** Pretension and **b.** prestress values for 2H-TaS$_2$ crystals of various thickness are shown.

**SEM Micrographs of AFM tips After a Single Measurement**

We characterized the changes on a single AFM tip using SEM imaging. We first acquired SEM image of a pristine tip. Then, scanned the sample and took another SEM image. We installed the same tip to the AFM, this time we scanned the sample and performed an indentation and took an SEM image. Finally, we used the same tip for a scan, indentation and breaking and acquired an SEM image. **Figure S9** shows the series of SEM images after each step and gradual degradation of the AFM tip.

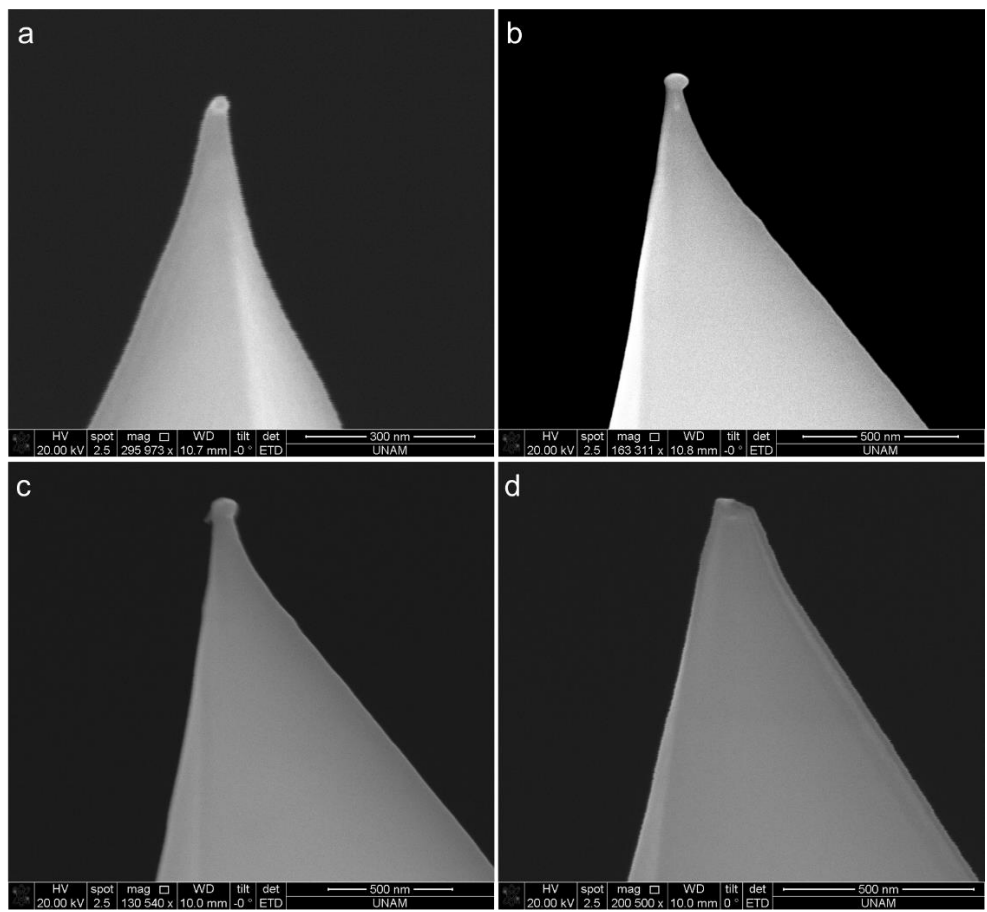

**Figure S9 a.** SEM micrograph of a pristine tip. The tip radius is measured as 12 nm. **b.** SEM micrograph of the same tip in **a** after a single scan typically performed to map the crystal. Tip radius is 23 nm. **c.** Same tip in **b** after a second scan followed by an indentation. Tip radius is 28 nm. **d.** Same tip in **c** after a third scan, indentation and a breaking. Tip is now blunt.

Before the measurements we picked up 10 random tips from the box and measured the diameter of the AFM tips. We found out that the tip radii are within the specifications of the manufacturer (10.0 ± 1.0 nm). We used each tip for a single indentation and breaking measurement and after every other measurement we measured the tip radius using SEM imaging (**Figure S10**). For the measurements we have the measured tip value, we used the measured value, otherwise we used the averaged tip radius 22.0 ± 2.0 nm.

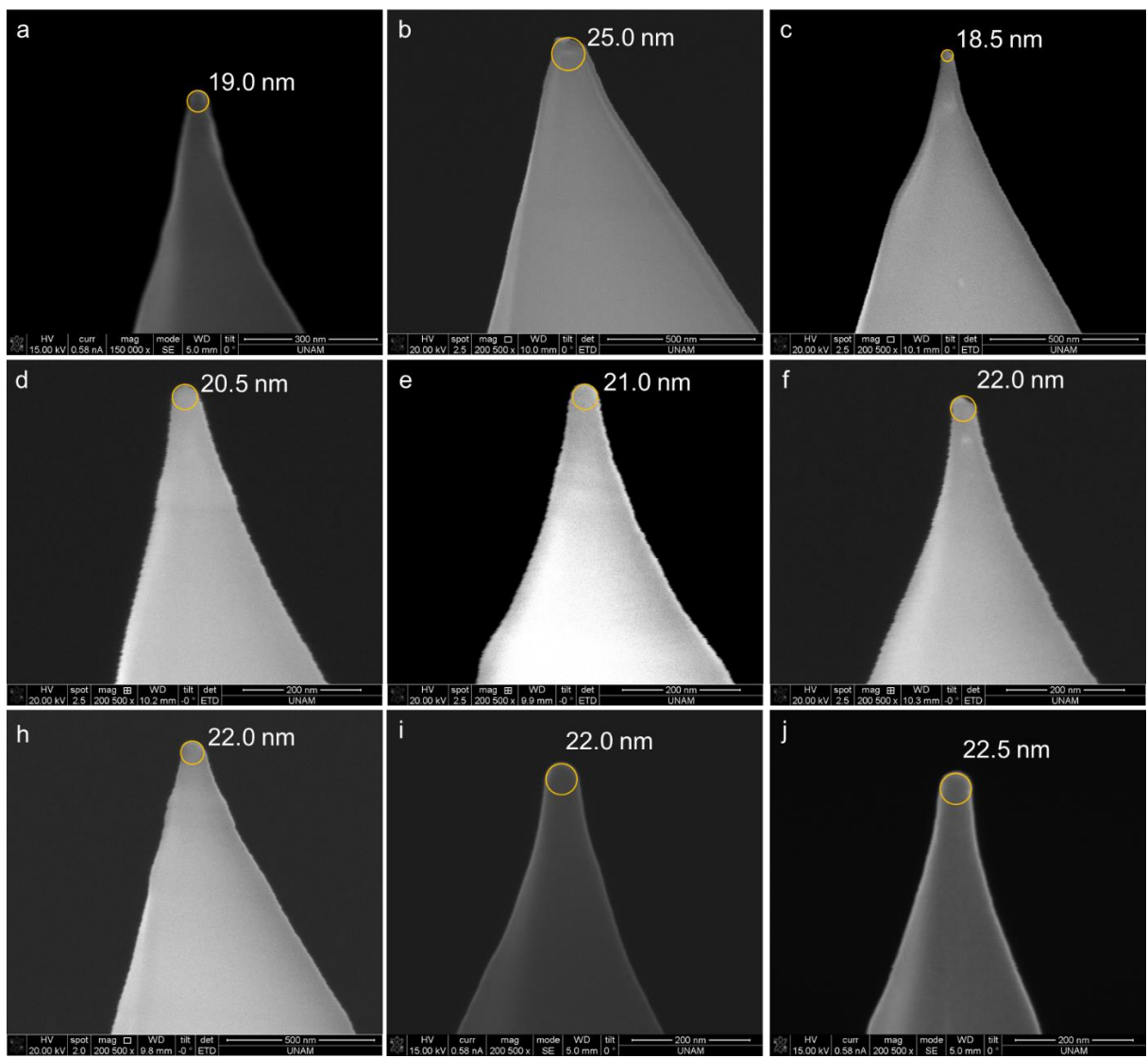

**Figure S10** SEM micrographs of DLC AFM tips after a single indentation and breaking measurement. The yellow circles show the tip apex, and the values are the tip radii.

**Crystal Structures of 2H-TaS$_2$, 1T-TaTe$_2$, 3R-NbS$_2$, 1T-NbTe$_2$**

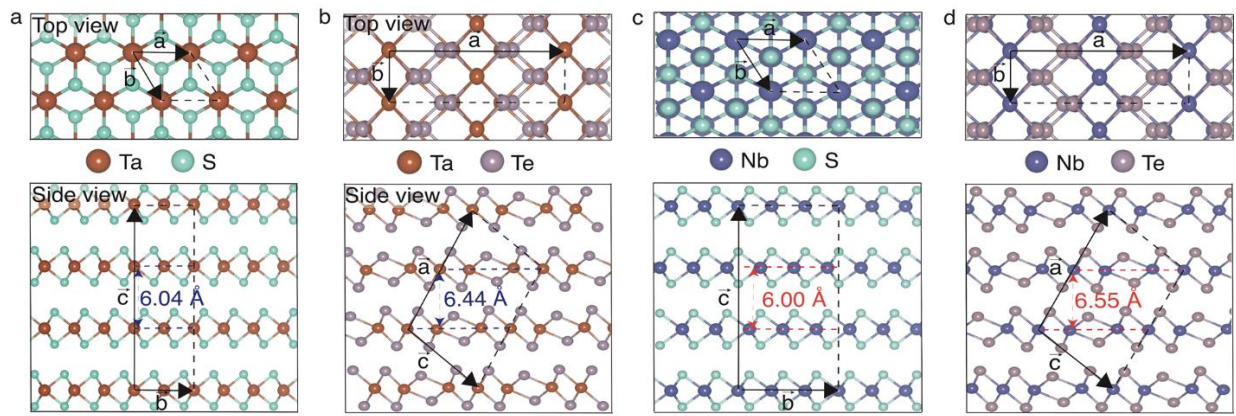

**Figure S11 a.** 2H-TaS$_2$ **b.** 1T-TaTe$_2$ **c.** 3R-NbS$_2$ **d.** 1T-NbTe$_2$

**XPS Survey on bulk metallic TMDCs**

We performed XPS surveys on exfoliated TDMCs right after exfoliation and after keeping the samples under ambient for an hour. Except TaS$_2$ other metallic TMDCs showed significant signs of oxidation. **Figure S12** shows XPS surveys of the chalcogen and metal constituents of the metallic TMDCs.

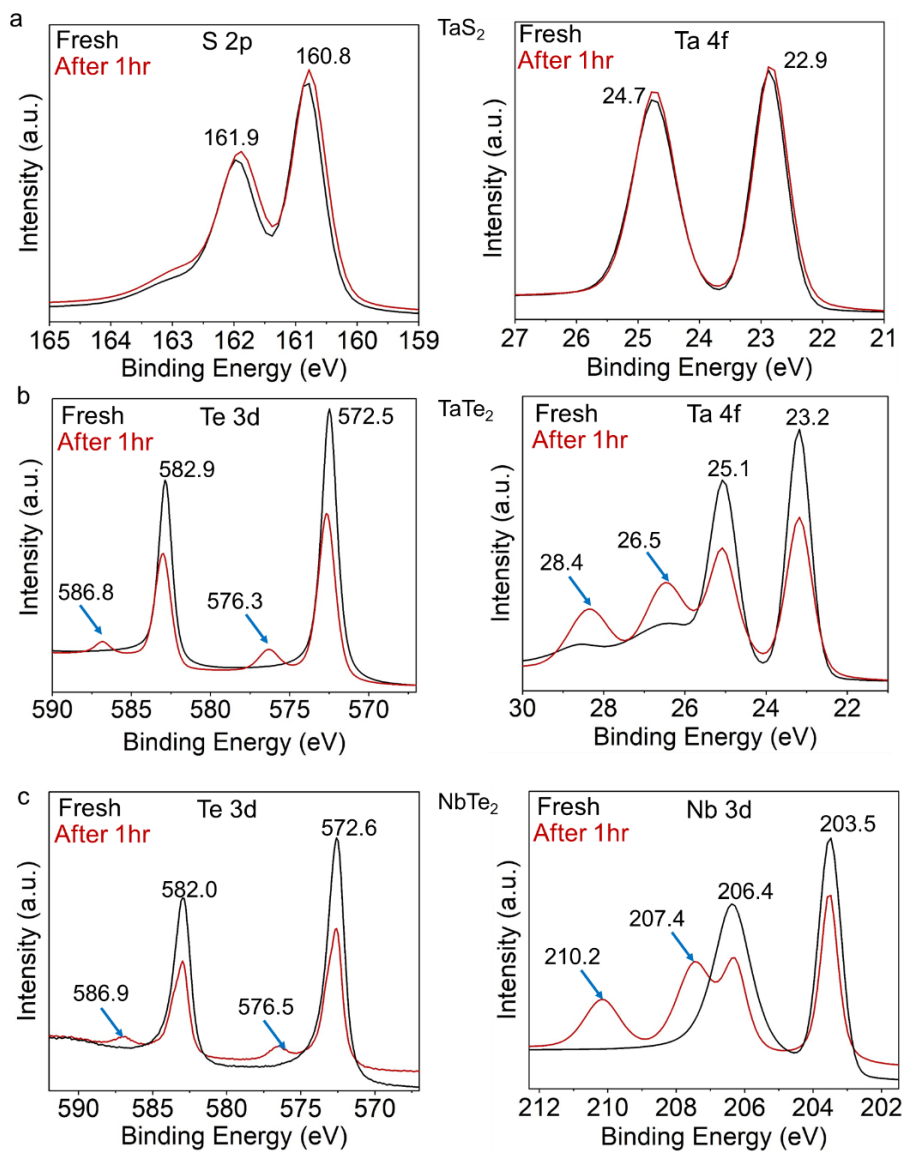

**Figure S12 a.** S 2p and Ta 4f surveys of $TaS_2$ taken on freshly exfoliated samples and after exposing to ambient for 1 hour. **b.** Te 3d and Ta 4f surveys of $TaTe_2$ taken on freshly exfoliated samples and after exposing to ambient for 1 hour. **c.** Te 3d and Nb 3d surveys of $NbTe_2$ taken on freshly exfoliated samples and after exposing to ambient for 1 hour. $TaTe_2$ and $NbTe_2$ show clear signs of oxidation after 1 hour in ambient conditions.

**References**

(1) Frank, I. W.; Tanenbaum, D. M.; van der Zande, A. M.; McEuen, P. L. Mechanical Properties of Suspended Graphene Sheets. *J. Vac. Sci. Technol. B Microelectron. Nanom. Struct.* **2007**, *25* (6), 2558. https://doi.org/10.1116/1.2789446.

(2) Hohenberg, P.; Kohn, W. Inhomogeneous Electron Gas. *Phys. Rev.* **1964**, *136* (3B), B864–B871. https://doi.org/10.1103/PhysRev.136.B864.

(3) Kohn, W.; Sham, L. J. Self-Consistent Equations Including Exchange and Correlation Effects. *Phys. Rev.* **1965**, *140* (4A), A1133–A1138.

https://doi.org/10.1103/PhysRev.140.A1133.

(4) Kresse, G.; Furthmüller, J. Efficiency of Ab-Initio Total Energy Calculations for Metals and Semiconductors Using a Plane-Wave Basis Set. *Comput. Mater. Sci.* **1996**, *6* (1), 15–50. https://doi.org/10.1016/0927-0256(96)00008-0.

(5) Blöchl, P. E. Projector Augmented-Wave Method. *Phys. Rev. B* **1994**, *50* (24), 17953–17979. https://doi.org/10.1103/PhysRevB.50.17953.

(6) Perdew, J. P.; Burke, K.; Ernzerhof, M. Generalized Gradient Approximation Made Simple. *Phys. Rev. Lett.* **1996**, *77* (18), 3865–3868. https://doi.org/10.1103/PhysRevLett.77.3865.

(7) Henkelman, G.; Arnaldsson, A.; Jónsson, H. A Fast and Robust Algorithm for Bader Decomposition of Charge Density. *Comput. Mater. Sci.* **2006**, *36* (3), 354–360. https://doi.org/10.1016/j.commatsci.2005.04.010.

(8) Kecik, D.; Onen, A.; Konuk, M.; Gürbüz, E.; Ersan, F.; Cahangirov, S.; Aktürk, E.; Durgun, E.; Ciraci, S. Fundamentals, Progress, and Future Directions of Nitride-Based Semiconductors and Their Composites in Two-Dimensional Limit: A First-Principles Perspective to Recent Synthesis. *Appl. Phys. Rev.* **2018**, *5* (1), 011105. https://doi.org/10.1063/1.4990377.

(9) Singh, S.; Valencia-Jaime, I.; Pavlic, O.; Romero, A. H. Elastic, Mechanical, and Thermodynamic Properties of Bi-Sb Binaries: Effect of Spin-Orbit Coupling. *Phys. Rev. B* **2018**, *97* (5), 054108. https://doi.org/10.1103/PhysRevB.97.054108.

(10) Hill, R. The Elastic Behaviour of a Crystalline Aggregate. *Proc. Phys. Soc. Sect. A* **1952**, *65* (5), 349–354. https://doi.org/10.1088/0370-1298/65/5/307.

(11) Nakashima, S.; Tokuda, Y.; Mitsuishi, A.; Aoki, R.; Hamaue, Y. Raman Scattering from 2H-NbS2 and Intercalated NbS2. *Solid State Commun.* **1982**, *42* (8), 601–604. https://doi.org/10.1016/0038-1098(82)90617-2.

(12) Hangyo, M.; Nakashima, S.-I.; Mitsuishi, A. Raman Spectroscopic Studies of MX2-Type Layered Compounds. *Ferroelectrics* **1983**, *52* (1), 151–159. https://doi.org/10.1080/00150198308208248.

(13) Sader, J. E.; Sanelli, J. A.; Adamson, B. D.; Monty, J. P.; Wei, X.; Crawford, S. A.; Friend, J. R.; Marusic, I.; Mulvaney, P.; Bieske, E. J. Spring Constant Calibration of Atomic Force Microscope Cantilevers of Arbitrary Shape. *Rev. Sci. Instrum.* **2012**, *83* (10). https://doi.org/10.1063/1.4757398.